\documentclass[sigconf]{acmart}

\usepackage{acmart-taps}
\usepackage{graphicx}
\usepackage{bm}
\usepackage{url}
\usepackage{color}
\usepackage{pgfplots}
\usepackage{pgfplotstable}
\usepackage{booktabs}
\usepackage{listings}
\usepackage{xcolor}
\usepackage{tablefootnote}
\usepackage{ulem}
\usepackage[font=footnotesize]{caption}

\newcommand{\gpluto}{\texttt{gPLUTO}}

\usepackage{cfcompanion26-28}

\begin{document}

\title{On the Limits of Performance Portability in Directive-Based GPU Programming}
\thanks{Pre-print version of the manuscript submitted to The 23rd ACM International Conference on Computing Frontiers (CF'26).}
\author{Alessandro Romeo}
\affiliation{%
  \department{SuperComputing Applications and Innovation Department}
  \institution{CINECA}
  \streetaddress{Via Magnanelli 6/3}
  \city{Casalecchio di Reno}
  \state{BO}
  \postcode{40033}
  \country{Italy}
}
\email{a.romeo@cineca.it}

\author{Nitin Shukla}
\affiliation{%
  \department{SuperComputing Applications and Innovation Department}
  \institution{CINECA}
  \streetaddress{Via Magnanelli 6/3}
  \city{Casalecchio di Reno}
  \state{BO}
  \postcode{40033}
  \country{Italy}
}

\author{Stefano Truzzi}
\affiliation{%
  \department{Dipartimento di Fisica}
  \institution{Università degli Studi di Torino}
  \streetaddress{Via Pietro Giuria 1}
  \city{Torino}
  \postcode{I-10125}
  \country{Italy}
}

\author{Alessio Suriano}
\affiliation{%
  \department{Dipartimento di Fisica}
  \institution{Università degli Studi di Torino}
  \streetaddress{Via Pietro Giuria 1}
  \city{Torino}
  \postcode{I-10125}
  \country{Italy}
}

\author{Andrea Mignone}
\affiliation{%
  \department{Dipartimento di Fisica}
  \institution{Università degli Studi di Torino}
  \streetaddress{Via Pietro Giuria 1}
  \city{Torino}
  \postcode{I-10125}
  \country{Italy}
}

\begin{abstract}
The transition of scientific applications to GPU-accelerated exascale systems is constrained by trade-offs between performance, portability, and productivity. This work evaluates the performance portability of directive-based GPU programming by porting \gpluto{}, a production-grade magnetohydrodynamics code for astrophysical simulations, from OpenACC to OpenMP, and analyzing its performance on NVIDIA A100 (Leonardo Booster) and AMD MI250X (LUMI-G) devices. On NVIDIA platforms, OpenACC and OpenMP achieve comparable performance due to a shared compiler backend, providing a consistent baseline for assessing algorithmic efficiency. In contrast, the same OpenMP implementation is approximately three times slower at the application level on AMD MI250X with respect to the NVIDIA A100 OpenACC baseline, with kernel-level slowdowns reaching up to an order of magnitude, driven by sensitivity to strided memory-access patterns and compiler limitations. Kernel-level profiling shows that the dominant contributors to runtime are memory-latency-bound rather than limited by peak bandwidth. In low-parallelism kernels, C++ abstraction layers increase register pressure and spilling, leading to extreme slowdowns of up to $47 \times$ in specific cases. These results indicate that portable performance across GPU architectures requires not only application-level changes but also continued advances in compiler backends and architecture-aware optimization strategies.
\end{abstract}

\keywords{GPU offloading, OpenMP, OpenACC, performance portability, High performance computing, \gpluto{}, GPU benchmarking, performance optimization,
heterogeneous systems, NVIDIA A100, AMD MI250X, Leonardo, LUMI}

\maketitle

\section{Introduction}
\label{sec:intro}
High-performance computing systems have become increasingly heterogeneous, with GPUs driving progress toward exascale computing ($10^{18}$ calculations per second): over 50\% of TOP500 systems now integrate accelerators to maximize throughput and energy efficiency~\footnote{\url{https://www.top500.org/lists/top500/2025/11/}}. While GPUs have opened new frontiers in scientific simulations, legacy applications face significant portability challenges, reflecting the long-standing trade-off between performance, portability, and productivity~\citep{Owens08,Edwards2013,Heroux2020}.
%High-performance computing systems have become increasingly heterogeneous by incorporating parallel processing units such as GPUs and FPGAs. GPU architectures with high-speed interconnects are driving progress toward exascale computing ($10^{18}$ calculations per second). Today, over 50\% of TOP500 supercomputers integrate GPUs or other accelerators to maximize throughput and energy efficiency~\footnote{\url{https://www.top500.org/lists/top500/2025/11/}}. GPUs have opened new frontiers in scientific simulations. Researchers explore complex physical systems at unprecedented resolutions and speeds. However, legacy applications face challenges running efficiently across different vendor platforms. The accelerator-based heterogeneity of modern exascale systems creates significant obstacles. This reflects the long-standing trade-off between performance, portability, and productivity~\citep{Owens08,Edwards2013,Heroux2020}.
Several EuroHPC Centers of Excellence across scientific domains - including SPACE\footnote{\url{https://www.space-coe.eu/}}~\citep{Shukla26,Shukla2025_EuroHPCDay}, Plasma-PEPSC\footnote{\url{https://plasma-pepsc.eu/}}~\citep{WilliamsetalPlasmaPEPSC2025}, MaX\footnote{\url{https://www.max-centre.eu/}}~\citep{MaX2025ComputingFrontiers}, and ChEESE-2P\footnote{\url{https://www.cheese2.eu/}}~\citep{CHEESEexascale2023} - support re-engineering flagship codes for exascale performance and portable HPC best practices. Transitioning large legacy scientific codes to GPUs raises fundamental questions: (i) Can they be ported with minimal effort? (ii) How can reproducibility be preserved across evolving hardware? (iii) How can performance portability be achieved without sacrificing efficiency?

Exploiting modern GPU memory bandwidths (2-3~TB/s) requires careful parallelism exposure and memory-access management. Low-level APIs (\texttt{CUDA}, \texttt{HIP}) promise peak performance but lack portability. C++ abstraction libraries (\texttt{SYCL}, \texttt{RAJA}, \texttt{Kokkos}) provide portability through templates but require extensive code modifications~\cite{Davis25}. Directive-based models (\texttt{OpenACC}~\cite{OpenACC2023}, \texttt{OpenMP}~\cite{OpenMP2021}) improve portability by abstracting device details, though often with performance penalties~\cite{Memeti17,Khalilov_2021,krishnasamy26}: OpenACC benefits from a mature NVIDIA CUDA backend, while OpenMP offloading shows uneven optimization across compilers and architectures~\cite{Antao2016,Deakin19}. While quantitative portability metrics exist~\cite{pennycook2016,Sewall20,Holmen2019,marowka25}, applying them to multi-kernel production scientific applications remains challenging. 

This work focuses on directive-based models due to their widespread adoption in legacy MHD-Godunov~\citep{stone2020athena,Grete2021} and particle-in-cell~\cite{Sishtla19,Myers21} codes. Large-scale astrophysical and plasma codes exemplify how memory latency, compiler behavior, and data layout interact to determine achievable performance on modern accelerators. In particular, this paper examines performance portability limitations of directive-based GPU codes by analyzing the porting of \gpluto{}~\footnote{https://plutocode.ph.unito.it/}, a  production-grade MHD code for astrophysical plasma simulations, from OpenACC to OpenMP. We hypothesize that such limitations arise primarily from compiler implementation differences rather than programming model semantics, though architectural characteristics (memory hierarchy organization and sensitivity to strided memory accesses) play a crucial role. To test this, we perform kernel-level analysis of throughput, arithmetic intensity, instruction-level parallelism, register usage, and cache efficiency across GPU architectures and compiler toolchains, exploring performance portability challenges and trade-offs for a real application memory-latency dominated~\cite{Wienke12,Deakin18,Tandon2024,Dubey2021,Mehta2021}.

The remainder of this paper is organized as follows. Section~\ref{sec:methodology} describes the methodology and performance metrics used throughout the study, as well as the practical challenges encountered during the porting process. Section~\ref{sec:results} presents application-level and kernel-level results for \gpluto{} on NVIDIA A100 and AMD MI250X GPUs. Section~\ref{sec:discussion} analyzes cross-architecture portability. Finally, Section~\ref{sec:conclusions} summarizes the main findings and discusses their implications for future GPU software development. This study operates under unified memory on both platforms (NVIDIA Managed Memory on A100, AMD Unified Memory on MI250X), as required for stable execution of \gpluto{}'s pointer-intensive C++ codebase. While this introduces vendor-specific runtime behaviors (page migration, fault handling), our analysis focuses on programming model semantics and compiler-generated code; disentangling memory management from compiler and architectural effects remains future work.
%This study operates under unified memory mechanisms on both platforms (NVIDIA Managed Memory on A100, AMD Unified Memory on MI250X), as required for stable execution of \gpluto{}'s pointer-intensive C++ codebase. While unified memory introduces vendor-specific runtime behaviors (page migration, fault handling), our analysis focuses on the performance impact of programming model semantics and compiler-generated code rather than on explicit data movement strategies. Disentangling the pure contribution of memory management mechanisms from compiler and architectural effects would require additional controlled experiments with explicit allocation APIs, which remains future work.
\section{Methodology and Experimental Setup}
\label{sec:methodology}
This section describes the experimental methodology adopted in this work. We present the targeted HPC clusters, as well as the production application \gpluto{}, the adopted strategies implemented to optimize its OpenMP translation from OpenACC, the profiling tools, and the performance and execution metrics employed throughout the analysis. 

\subsection{Platform Details and Compilation Settings}
Experiments were conducted on NVIDIA A100 (Leonardo Booster) and AMD MI250X (LUMI-G) using single CPU-GPU nodes to isolate accelerator performance and eliminate inter-node communication effects. 
\label{sec:platformdet}
\begin{table}[h]
\centering
\scriptsize
\caption{GPU architecture specifications: Leonardo Booster (NVIDIA A100) and LUMI-G (AMD MI250X, single GCD). Work-items $\equiv$ NVIDIA threads.}
\begin{tabular}{|l|c|c|}
\hline
\textbf{Specification} & \textbf{A100} & \textbf{MI250X (1 GCD)} \\
\hline
Compute Units & 108 SMs & 110 CUs \\
Peak FP64 (TFLOP/s) & 9.7 & 13.3 (26.5 full MCM) \\
Max Base Clock (GHz) & 1.41 & 1.70 \\
HBM Bandwidth (TB/s) & 2.0 & 1.6 (3.2 full MCM) \\
HBM Capacity (GB) & 64 & 64 (128 full MCM) \\
L1 Cache (KB) & 192 & 64 \\
L2 Cache (MB) & 40 & 8 \\
Max 32-bit registers & 255 (per thread) & 256 (per work-item) \\
Vector register file [KB] & 256 (64K$\times$32-bit per SM) & 512 (per CU) \\
Execution group size & Warp = 32 threads & Wavefront = 64 threads \\
Register file scope & per SM & per CU \\
\hline
\end{tabular}
\label{tab:gpu_specs}
\end{table}
Table~\ref{tab:gpu_specs} summarizes the architectural differences. While the A100 features a monolithic design~\cite{choquette2021a100}, the MI250X is a multi-chip module (MCM) where each Graphics Compute Die (GCD) exposes distinct memory domains~\cite{Smith2022}. The MI250X offers higher peak FP64 throughput and larger register files, whereas the A100 provides significantly larger L2 capacity and a configurable L1/shared memory~\cite{ALCF_A100}. In particular, MI250X employs a disaggregated per‑CU front‑end with smaller caches (16~KB vector L1 per CU, plus 16~KB scalar L1 and 32~KB instruction L1 per 2~CUs)~\cite{ROCmSpecs}. Beyond an almost identical per-work-item cap (256 vs. 255), MI250X provisions a larger CU-level vector register file ($\approx$512~KB per CU) than A100 per SM (64K 32-bit regs $\equiv$ 256~KB per SM). Note that the register-file difference is independent of cache sizes~\cite{AmpereTuningGuide,ROCmSpecs}. These architectural variations, summarized in Table~\ref{tab:gpu_specs}, directly impact memory-latency tolerance and occupancy\footnote{\footnotesize Note that registers are allocated per thread/work-item (32-bit) while occupancy is limited by the SM/CU-level register file and by the execution group granularity.}. On Leonardo (NVIDIA A100), both the OpenACC and the OpenMP versions are compiled with the NVIDIA HPC SDK (\texttt{nvc++}) 24.5, using managed memory and relocatable device code. Optimization and inlining are enabled to match production settings. On LUMI we use \texttt{ROCm} 6.2.4 and the LLVM/Clang OpenMP offload toolchain\footnote{(i) \texttt{craype-x86-trento}, (ii) \texttt{PrgEnv-amd/8.5.0}, (iii) \texttt{craype-accel-amd-gfx90a}, and (iv) \texttt{rocm/6.2.4} (\texttt{clang++}).} provided by the system environment with \texttt{-O3} optimization. To ensure a fair comparison with the monolithic A100 and avoid variable NUMA placement effects, single-GPU runs on LUMI were explicitly bound to a specific GCD using \texttt{ROCR\_VISIBLE\_DEVICES}. This setting does not alter kernel execution behavior or enable additional optimizations. Both platforms utilized unified memory mechanisms (NVIDIA Managed Memory and AMD's \texttt{CRAY\_ACC\_USE\_UNIFIED\_MEM=1} and \texttt{HSA\_XNACK=1}) to support \gpluto{}'s pointer-intensive C++ structures. While transparent page migration may introduce performance overhead compared to explicit \texttt{target data} regions, these settings were necessary for code stability across both architectures.

\subsection{\gpluto{}: Production Astrophysics Code}
\label{sec:realworldapp}
\gpluto{} is a finite-volume, Godunov-type code solving gas and plasma dynamics equations in astrophysical contexts. It supports five physics modules ranging from classical hydrodynamics to resistive relativistic magnetohydrodynamics. 
\subsubsection{Multi-Vendor GPU Implementation of \gpluto{}}
\gpluto{} porting began in 2023 within SPACE-CoE~\cite{rossazza2025plutocodegpuslook}, offloading approximately 70\% of modules to GPUs with strong NVIDIA performance. The code was redesigned for exascale GPU computing, using OpenACC for acceleration and scaling efficiently to thousands of GPUs with non-blocking MPI and asynchronous data exchange. Development involved extensive parallelization optimization and a C-to-C++ rewrite introducing multidimensional array classes and templates. However, limited OpenACC compiler support for AMD C/C++ prompted an OpenMP adoption~\cite{suriano2025}. OpenMP directives were inserted manually or via Intel's automated translation framework\footnote{\url{https://github.com/intel/intel-application-migration-tool-for-openacc-to-openmp}}, though manual tuning remains necessary for performance-critical kernels. This approach preserves productivity while achieving vendor-neutral portability.

\subsubsection{Practical Lessons: OpenACC to OpenMP Porting}
In agreement with other works, achieving OpenACC-comparable OpenMP performance requires careful redesign beyond mechanical \texttt{pragma} replacement. Such a process involves refactoring, aliasing control, and careful handling of data abstractions~\cite{ALDINUCCI202113, Fridman25}. While OpenMP offloading has matured through multiple compiler generations~\cite{Bertolli2014, yan2025, deakin2023programming_openmp_gpu}, production deployment still requires detailed profiling and kernel-specific tuning. For the work within \gpluto{}, key findings include:

\textbf{Parallelism exposure and nested loop restructuring:}
although OpenMP 5.x has relaxed perfect-nesting constraints, GPU offloading often exposes only two reliable parallelism levels (\texttt{teams} and \texttt{threads}), as \texttt{simd} support and its mapping to GPU lanes remains compiler-dependent. Optimal directives (\texttt{teams loop} vs. \texttt{teams distribute parallel for}) vary by kernel, compiler (NVHPC vs. ROCm), and architecture. Achieving full occupancy frequently requires \texttt{collapse(3)} on spatial loop nests, necessitating code restructuring: intermediate declarations must move inside innermost loops, improving exposed parallelism but increasing register pressure through repeated initialization and reduced compiler hoisting. In contrast, OpenACC's \texttt{gang/worker/vector} hierarchy maps iteration spaces without full collapsing, allowing intermediate declarations outside vector loops and reducing register pressure.

\begin{imageonly}
\begin{Ccode}
#pragma omp target teams distribute parallel for collapse(2)
#pragma acc parallel loop collapse(2)
for (k = kbeg; k <= kend; k++){
  for (j = jbeg; j <= jend; j++){
    // Intermediate declarations / temporaries not depending on i
    // (can stay here in OpenACC thanks to vector level)
    ...
    // No guaranteed omp simd mapping across compilers
    #pragma acc loop vector
    for (i = ibeg; i <= iend; i++){
      ...
    }  }  }
\end{Ccode}
\end{imageonly}

\noindent Optimizations improving OpenMP performance may therefore degrade it through register pressure, while OpenACC achieves equivalent parallelism without restructuring: optimizing for one model may produce a performance loss for the other. Moreover, non-perfect loop collapsing availability remains limited and compiler-dependent on EuroHPC clusters.

\textbf{Data management and C++ abstraction handling:} while OpenMP 5.x supports implicit mapping and unified shared memory, optimal performance for pointer-based C++ structures requires explicit \texttt{map} clauses or \texttt{declare mapper} directives. Without them, compilers generate redundant private copies, increasing register pressure and spilling. OpenACC handles such abstractions more robustly via implicit deep-copy semantics. For instance, OpenACC's \texttt{present} clause enforces device residency with runtime errors; OpenMP's \texttt{present} modifier silently creates new mappings if data is absent, risking unexpected transfers. Complex structures require \texttt{declare mapper} or pointer APIs (e.g., \texttt{use\_device\_ptr}, \texttt{omp\_target\_alloc}), increasing complexity. Incorrect mapper definitions can cause silent data corruption or device segmentation faults, with limited compiler diagnostics to aid debugging. Additionally, \texttt{mapper} support varies across compilers, and complex nested structures or templated wrappers may prevent successful instantiation, increasing maintenance burden and significant code modifications.
%while OpenMP 5.x supports implicit mapping and unified shared memory, achieving optimal performance for complex pointer-based C++ structures often requires explicit \texttt{map} clauses or \texttt{declare mapper} directives in practice. Without it, compilers conservatively generate redundant private copies, increasing register pressure and triggering spilling. In contrast, OpenACC handles such abstractions more robustly through implicit deep-copy semantics and dedicated directives. For instance, OpenACC's \texttt{present} clause enforces device residency with runtime errors; OpenMP's \texttt{present} modifier silently creates new mappings if data is absent, risking unexpected transfers. Complex structures require \texttt{declare mapper} or pointer APIs (\texttt{use\_device\_ptr}, \texttt{omp\_target\_alloc}), increasing complexity. Incorrect mapper definitions can cause silent data corruption or device segmentation faults, with limited compiler diagnostics to aid debugging. Unified memory mechanisms further complicate OpenMP data regions through transparent page migration, reducing programmer control over placement. Furthermore, \texttt{mapper} support varies across compilers (NVHPC, Clang, GCC, AOCC), and complex nested structures may prevent successful mapper instantiation. Templated wrappers, common in modern C++ scientific codes, require explicit template instantiation per mapper, increasing code maintenance burden.

\textbf{Compiler-specific issues:} 
they include NVHPC 24.5 internal errors with separated \texttt{teams distribute + parallel for} (resolved in 24.7), incomplete \texttt{simd} support in ROCm 6.2.4, and subtle differences in reduction semantics, sequential loop enforcement and device function declarations.
%NVHPC 24.5 showed internal errors with separated \texttt{teams distribute + parallel for} (resolved in 24.7). ROCm 6.2.4 lacks full \texttt{simd} support for GPU vectorization. Reduction semantics, device functions (\texttt{acc routine} vs. \texttt{omp declare target}), and sequential loop enforcement differ subtly, requiring per-compiler validation.

\subsubsection{\gpluto{}'s main cycle and adopted tests}
Concerning \gpluto{} behavior, the code evolves conservation laws through three key stages — reconstruction of primitive (or characteristic)  variables, Riemann problem solution at cell interfaces, and flux-conservative temporal update — executed as distinct GPU kernels repeated per spatial dimension. Additional procedures are required in the presence of magnetic fields (e.g., constrained transport routines) and for Runge-Kutta ($3^{\rm rd}$-order by default) time integration, as well as for setting boundary conditions and performing domain exchanges through MPI communication. We focus on the 3D Orszag–Tang vortex problem ($352^3$ grid) and the 2D Riemann test, profiling kernels \texttt{Reconstruct()}, \texttt{HLLD\_Solver()} (or \texttt{HLL\_Solver()} for Riemann 2D), \texttt{RightHandSide()}, and auxiliary routines (\texttt{CT\_EMF()}, \texttt{US()}).

The 3D Orszag–Tang problem with $352^3$ grid ($\approx 43.6$ million cells) was selected to represent realistic \gpluto{} production workloads while ensuring adequate GPU utilization. This configuration generates approximately 170,000 thread blocks ($\lceil 352^3/256 \rceil$), providing 1,500–1,600 blocks 
per SM/CU and exceeding hardware concurrency limits.
%\footnote{Calculated as $\lceil 352^3 / \text{block\_size} \rceil$ with typical block size of 256 threads, yielding $\lceil 43,614,208 / 256 \rceil = 170,368$ blocks.}, providing 1,500–1,600 blocks per SM/CU, far exceeding the hardware concurrency limits (16–32 concurrent blocks per SM on A100; 40 wavefronts per CU on MI250X). 
Profiling confirms that observed occupancy limitations (12–71\%, Tables~\ref{tab:leonardo_merged} and~\ref{tab:lumi_kernels}) originate from register pressure and memory-latency effects rather than insufficient exposed parallelism. The 4.3$\times$ higher parallelism relative to the 2D Riemann case ($3200^2 \approx 10.2$ million cells) enables both OpenACC and OpenMP to effectively hide register-spilling latency in 3D.

%\subsection{Profiling Tools}
%\label{sec:proftools}
\subsection{Profiling Tools and Performance Metrics}
\label{sec:all_metrics}
Performance analysis employed NVIDIA Nsight Compute/Systems for A100 and AMD rocProfiler/OmniPerf for MI250X. These tools collect hardware counters for throughput, bandwidth, cache efficiency, and instruction-level metrics following vendor roofline methodologies~\cite{nvidia_nsight_roofline, amd_omniperf, ENCCS22_Hierarchical_Roofline}.

Metrics are derived from architecture-specific hardware counters. These metrics are functionally analogous across platforms but not numerically identical; cross-architecture comparisons therefore identify execution regimes qualitatively rather than comparing absolute values. We report achieved metrics (occupancy, bandwidth) measured at runtime rather than theoretical peaks, as they reflect real kernel execution under register pressure, memory latency, and instruction scheduling.

\textbf{Kernel runtimes (ms) and fraction per step (\%):}
kernel execution times on NVIDIA A100 are obtained from Nsight Systems, while on AMD MI250X they are obtained from ROCm profiling tools (RocProf/OmniPerf). Reported single-kernel times correspond to steady-state execution and may slightly overestimate per-step costs due to initialization and profiling overheads. Only the dominant computational kernels are included in the runtime breakdown; auxiliary kernels (e.g., boundary conditions) account for the remaining execution time.
The execution times reported in Section~\ref{sec:results} refer to a single kernel invocation (e.g., one directional sweep for one Runge-Kutta stage) measured under isolated profiling conditions. 

Runtime fractions are computed per Runge-Kutta time step by aggregating all directional variants and stage repetitions of each kernel. Fractions are derived from application runs with profiling enabled; while absolute timings may be affected by profiling overhead, relative kernel contributions are preserved. The same methodology is applied consistently to both Leonardo and LUMI results.

\textbf{Computational Throughput (GFLOP/s):} this metric quantifies sustained 
double-precision floating-point (FP64) operations per second:
\[\small
\text {FP64 Computational Throughput} = \frac{\text{FP64 FLOPs}}{\text{Runtime (s)} \times 10^9}.
\]
It is derived from hardware counters normalized for vector width on AMD. VALU utilization on MI250X indicates instruction issue cycles, not peak throughput, making FP64 throughput the preferred metric. This choice is further justified a posteriori, as all examined \gpluto{} kernels predominantly execute FP64 arithmetic operations.

\textbf{Achieved Memory Bandwidth (GB/s):} data transferred per second between L2 cache and high-bandwidth memory (HBM), capturing DRAM traffic:
\[\small
\text{Memory Bandwidth} = \frac{\text{Bytes Transferred}}{\text{Runtime (s)} \times 10^9}.
\]
This quantifies HBM pressure following the aforementioned hierarchical roofline methodology. High L2 hit rates reduce HBM bandwidth without reducing memory activity.

\textbf{Arithmetic Intensity (AI):} FP64 FLOPs per HBM byte transferred:
\[\small
\text{AI} = \frac{\text{FP64 FLOPs}}{\text{Bytes Transferred}}.
\]
HBM-based AI enables consistent identification of compute- versus memory-limited regimes across architectures~\citep{williams2009roofline, choquette2021a100, schieffer2025} without reconstructing pre-cache traffic, which is architecture-dependent and not directly comparable.

\textbf{Additional Metrics:} instructions per cycle (IPC, due to fundamental differences in execution models, used for intra-architecture analysis only); register usage (measuring spilling probability into local memory); achieved occupancy (the fraction of active warps or wavefronts relative to hardware limits, reflecting kernels' ability to hide latency through concurrency); L1/L2 cache hit rates (to estimate spatial locality); branch utilization and efficiency (to quantify control-flow divergence within warps or wavefronts); memory-latency indicators.

\textbf{Kernel Classification:} kernels are classified as \emph{compute-bound} (high AI, high throughput), \emph{memory-bandwidth-bound} (low AI, high achieved bandwidth), \emph{memory-latency-bound} (low bandwidth despite frequent operations, low utilization), or \emph{control-flow-bound} (branch-dominated with low AI, low execution-unit utilization, and limited memory throughput). Empirically a compute-bound kernel exhibits AI $>$ 5~flop/byte and utilization $>$ 30\%; latency-bound exhibits AI $<$ 2~flop/byte and bandwidth $<$ 20\% of peak.
\section{Results}
\label{sec:results}
In the following sections OpenACC on NVHPC represents the most mature implementation and establishes performance upper bounds. Kernel metrics represent median or average values from steady-state time steps, excluding initialization 
overheads. Cross-architecture comparisons use the same OpenMP implementation to isolate architectural effects.
\begin{table*}[htbp]
\centering
\scriptsize
\caption{\gpluto{} kernel metrics on NVIDIA A100 (\texttt{NVHPC} 24.5), 3D Orszag-Tang. OpenACC and OpenMP achieve near-identical performance (except \texttt{CT\_EMF()}) via the unified CUDA backend.}
\label{tab:leonardo_merged}
\begin{tabular}{|l|cc|cc|cc|cc|cc|}
\hline
{\textbf{Metric}} & 
\multicolumn{2}{c|}{\textbf{\texttt{HLLD()}}} & 
\multicolumn{2}{c|}{\textbf{\texttt{Reconstruct()}}} & 
\multicolumn{2}{c|}{\textbf{\texttt{RHS()}}} & 
\multicolumn{2}{c|}{\textbf{\texttt{CT\_EMF()}}} & 
\multicolumn{2}{c|}{\textbf{\texttt{US()}}} \\
\cline{2-11}
& ACC & OMP & ACC & OMP & ACC & OMP & ACC & OMP & ACC & OMP \\
\hline
Time (ms) & 31.6 & 32.0 & 13.7 & 14.6 & 9.0 & 9.3 & 7.8 & 23.9 & 18.8 & 16.6 \\
Runtime fraction (\%) & 20.5 & 16.7 & 8.9 & 7.6 & 5.8 & 4.9 & 10.1 & 25.0 & 12.2 & 8.7 \\
FP64 ThrP (GFlop/s) & 71.3 & 66.9 & 217.1 & 213.6 & 11.0 & 11.0 & 15.5 & 18.7 & 0.85 & 0.76 \\
Mem BW (GB/s) & 48.9 & 42.1 & 35.3 & 35.0 & 32.1 & 31.0 & 83 & 278 & 22.7 & 22.1 \\
AI (flop/byte) & 1.46 & 1.61 & 6.16 & 6.12 & 0.34 & 0.36 & 0.19 & 0.10 & 0.04 & 0.05 \\
Occupancy (\%) & 12 & 12 & 48 & 37 & 54 & 48 & 37 & 46 & 64 & 52 \\
IPC & 0.10 & 0.10 & 0.51 & 0.54 & 0.08 & 0.09 & 0.02 & 0.02 & 0.09 & 0.14 \\
L1 Hit (\%) & 50.4 & 58.1 & 62.7 & 62.7 & 57.5 & 49.7 & 18.3 & 18.5 & 23.5 & 26.6 \\
L2 Hit (\%) & 87.5 & 87.2 & 78.1 & 77.9 & 79.8 & 79.5 & 68.2 & 55.8 & 81.7 & 84.4 \\ 
Registers/thread & 172 & 182 & 64 & 74 & 56 & 64 & 80 & 65 & 38 & 56 \\
Spill Loads $\times10^{7}$ & 8.8 & 9.0 & 0 & 0 & 0 & 0 & 0 & 0 & 0 & 0 \\
Spill Stores $\times10^{7}$ & 4.9 & 5.3 & 0 & 0 & 0 & 0 & 0 & 0 & 0 & 0 \\
Branch Instr. $\times10^{7}$ & 15.8 & 0.4 & 37.4 & 10.8 & 1.8 & 0.1 & 0.9 & 0.14 & 12 & 15.2 \\
Avg Divergent Branches & 501 & 450 & 0 & 0 & 0 & 0 & 502 & 0 & 0 & 0 \\ 
Branch Efficiency (\%) & 99.7 & 99.8 & 100 & 100 & 100 & 100 & 95.5 & 100 & 100 & 100 \\
\hline
\end{tabular}
\end{table*}
 
%In the following sections we present key findings comparing OpenACC and OpenMP adoption alongside detailed performance analysis on HPC platforms described in Section~\ref{sec:platformdet}. OpenACC on NVHPC represents the most mature implementation and establishes an upper bound on expected performance. The identification and detailed steady-state analysis of performance-critical kernels has been assessed using NVIDIA Nsight Compute/Systems and AMD rocProfiler/OmniPerf. Kernel metrics were measured multiple times, with reported values representing median or average measurements. We report measurements from steady-state time steps, excluding the first iteration(s), which include one-time initialization overheads such as memory allocation, data migration, and runtime setup. Cross-architecture comparisons use the same OpenMP implementation to isolate architectural effects. 

\subsection{\gpluto{} Performance on Leonardo (NVIDIA A100)}
\label{sec:orszagtangLeo}

Table~\ref{tab:leonardo_merged} presents average metrics across directional kernel variants for the 3D Orszag-Tang case. Nsight Systems profiling reveals mild directional dependence: \texttt{HLLD()} $j$-sweeps execute approximately 20\% slower than $i,k$ sweeps on A100, while \texttt{RHS()} exhibits approximately 14\% directional variation. The \texttt{US()} kernels demonstrate factor-of-two runtime differences but maintain similar performance characteristics. Metrics are hence averaged across directional variants ($i/j/k$) for the same kernel type and across multiple profiling runs to reduce measurement noise.

\texttt{HLLD()} is primarily instruction-latency-bound: extreme register pressure ($>$ 170/thread) collapses occupancy to 12\% and triggers massive spilling (approximately $8.8 \times 10^7$ spill loads and $4.9 \times 10^7$ spill stores per launch), independently of the programming model used. This prevents the SM from hiding latency due to insufficient concurrent warps, despite high cache hit rates (L1 is approximately 50\% while L2 is 87\%). Low SM utilization ($<$ 1\%) and low throughput (approximately 50 GB/s versus 2~TB/s peak) decoupled from HBM traffic confirm that register-resident operations and latency, rather than bandwidth or memory locality, are the bottlenecks. In fact, HBM-based AI slightly underestimates sustained throughput. Even minor register increases disproportionately degrade performance, as SMs cannot hide instruction latency. Differences in branch instruction counts reflect compiler-level if-conversion and predication, but do not translate into runtime improvements due to the kernel being latency-bound. In contrast, \texttt{Reconstruct()} achieves balanced performance with moderate AI ($\approx$ 6~flop/byte), acceptable occupancy (37–48\%), and regular access patterns. L1 locality is high (approximately 63\%), and L2 reuse is also high (approximately 78\%). No register spilling occurs with 64–74 registers per thread. The near-identical performance between OpenACC and OpenMP confirms the efficiency of the unified CUDA backend for kernels with low register pressure. 
%\texttt{Reconstruct()} achieves the most balanced performance. Both OpenACC and OpenMP offloading achieve nearly identical performance on A100. This kernel exhibits moderate AI (approximately 6~flop/byte), acceptable occupancy (37–48\%), and nearly identical bandwidth under both programming models (approximately 35~GB/s). L1 locality is moderate (approximately 35\%), while L2 reuse is fair (approximately 63\%). No register spilling occurs with 64–74 registers per thread. The similar performance across programming models confirms the effectiveness of the unified CUDA backend when register pressure is low and memory access patterns are regular. 
\texttt{RHS()}, \texttt{CT\_EMF()}, and \texttt{US()} remain memory-latency-bound (AI < 0.4~flop/byte). Despite 37-64\% occupancy, all exhibit limited L1 locality (18–57\%) despite moderate-to-high L2 hit rates (56–84\%), minimal spilling, yet bandwidth 2-3 orders below peak. \texttt{CT\_EMF()} bandwidth variability reflects directional access anisotropy\footnote{This term describes performance variation when the same kernel accesses multidimensional arrays along different spatial dimensions.}, not improved computational efficiency. \texttt{CT\_EMF()} shows the largest OpenACC-OpenMP gap, suggesting its OpenMP mapping/privatization is suboptimal; kernel-specific tuning (loop restructuring, temporary placement) would likely improve parity\footnote{\texttt{CT\_EMF()} bandwidth varies directionally (OpenACC: 83~GB/s avg; OpenMP: 181–418~GB/s per sweep), yet execution time differs by $<$10\%, confirming latency-bound behavior.}. 

Overall, no kernels reach compute- or bandwidth-bound regimes. OpenMP exhibits a systematic 5-10\% slowdown due to slightly higher register pressure, necessitating iterative tuning to reach near-parity with OpenACC.

%Overall, no kernels are truly compute- or bandwidth-bound. \texttt{HLLD()} is latency-bound due to register pressure, while \texttt{US()} is latency-bound due to low arithmetic intensity. \texttt{RHS()} and \texttt{CT\_EMF()} are limited by irregular memory access. Reconstruction kernels represent the only region approaching balanced compute-memory behavior. However, even there SM throughput remains far from architectural limits. OpenMP exhibits a systematic 5–10\% slowdown versus OpenACC. This is attributed to slightly higher register pressure reducing occupancy rather than differences in instructions per cycle, which are nearly identical. Achieving this near-parity required iterative tuning. Mechanical OpenACC-to-OpenMP translation is insufficient, as demonstrated by the 2D Riemann case below.
\subsubsection{OpenMP Fragility in Low-Parallelism Regimes: 2D Riemann}
\label{sec:riemann}
The 2D Riemann test ($3200 \times 3200 \times 1$ domain, over $10^7$ iterations) reveals stark OpenACC-OpenMP differences absent in 3D configurations. Table~\ref{tab:riemann2D_kernels} shows that OpenMP is 47$\times$ slower for \texttt{Reconstruct()} and 6$\times$ slower for the \texttt{HLL()} Riemann Solver, despite a nominally large iteration space. 
\begin{table}[htbp]
\centering
\scriptsize
\caption{Riemann 2D kernel metrics on NVIDIA A100. OpenMP exhibits severe register spilling and suboptimal grid mapping despite large iteration space, causing order-of-magnitude slowdowns.}
\label{tab:riemann2D_kernels}
\begin{tabular}{|l|c|c|c|c|}
\hline
\textbf{Metric} & \textbf{\texttt{Rec\_ACC()}} & \textbf{\texttt{Rec\_OMP()}} & \textbf{\texttt{HLL\_ACC()}} & \textbf{\texttt{HLL\_OMP()}} \\
\hline
Grid size & 3200$\times$128 & 1257$\times$128 & 3200$\times$128 & 3200$\times$128 \\
Time (ms) & 1.24 & 58 & 1.63 & 9.84 \\
FP64 ThrP (GFlop/s) & 17.9 & 24.55 & 15.06 & 27.73 \\ 
Mem BW (GB/s) & 1003 & 240 & 1000 & 505 \\
Memory Throughput (\%) & 63.03 & 14.76 & 61.24 & 31 \\
AI (flop/byte) & 1.29 & 0.15 & 1.07 & 0.38 \\
Occupancy (\%) & 20.14 & 24.25 & 30.28 & 33.8 \\
L1 Hit (\%) & 66 & 61 & 10.87 & 66.5 \\
L2 Hit (\%) & 76 & 88 & 66 & 79 \\
Registers/thread & 104 & 98 & 90 & 80 \\
Spill Loads $\times10^{7}$ & 0 & 8.8 & 0 & 4.4 \\ 
Spill Stores $\times10^{7}$ & 0 & 5.5 & 0 & 1.1 \\
Branch Instr. $\times10^{7}$ & 1.2 & 8.8 & 0.69 & 1.8 \\
Avg Divergent Branches & 67.82 & 119.5 & 19.35 & 19.35 \\ 
Branch Efficiency (\%) & 99.53 & 99.92 & 99.77 & 99.93 \\
\hline
\end{tabular}
\end{table}
Unlike many small 2D test cases reported in the literature~\citep{deakin2023programming_openmp_gpu,klemm2025_openmp_offload_tutorial,Wienke14}, this poor performance cannot be attributed to insufficient parallelism. Two root causes explain this behavior. First, register spilling: OpenMP generates $8.8 \times 10^7$ spill loads for \texttt{Reconstruct()} (compared with zero for OpenACC), dominating execution. This spilling is driven by conservative privatization of C++ wrapper objects (\texttt{Ary1D}/\texttt{Ary2D}) without using mappers rather than by the raw register count, which is actually comparable across models. Second, suboptimal grid mapping: NVHPC OpenMP selects $1257\times 128$ grid vs. $3200\times128$ for OpenACC, exposing 2.5$\times$ less parallelism and reducing latency hiding capability. This compiler-driven grid selection does not reflect OpenMP model limitations; enforcing larger grids requires \texttt{num\_teams} or \texttt{thread\_limit} clauses, introducing machine-specific tuning that undermines portability. The trivial ($k=1$) dimension exacerbates this; OpenACC handles it transparently through automatic \texttt{collapse}, while OpenMP exhibits sensitivity to local object placement and nested loop structure. 

To quantify this sensitivity, we considered the best-case OpenMP configuration, analogous to current OpenACC implementation, which maps iteration space to GPUs. This is shown in the following snippet:

\begin{imageonly}
\begin{Ccode}
#pragma acc parallel loop collapse(2) present(d, Dts, grid)
#pragma omp target teams loop collapse(2) 
for (k = kbeg; k <= kend; k++){
  for (j = jbeg; j <= jend; j++){
    long int offset  = ni*(j + nj*k);
    Ary1D cmax(&Dts->cmax[offset],ni);
    ...
    Ary2D vL(&d->sweep.vL[offset1],ni,NVAR);
    ...
    #pragma acc loop
    #pragma omp loop
    for (i = ibeg; i <= iend; i++){
       ...
    } } }
\end{Ccode}
\end{imageonly}

\noindent and was compared against three alternative offloading strategies: (i) \texttt{omp target teams loop collapse(3)}, with all declarations inside the \texttt{i}-loop; (ii) \texttt{omp target teams distribute parallel for collapse(3)}, with all declarations moved inside the innermost \texttt{i}-loop; (iii) \texttt{omp target teams distribute parallel for collapse(2)}, leaving declarations between the \texttt{i}- and \texttt{j}-loops and applying \texttt{omp simd} before the \texttt{i}-loop. 
\begin{table}[H]
\centering
\scriptsize
\caption{OpenMP directive sensitivity for \texttt{HLL()} (Riemann 2D): minor syntax changes cause up to 22$\times$ runtime variation.}
\label{tab:HLL_OMP_variants}
\begin{tabular}{|l|c|c|c|c|}
\hline
\textbf{Metric} & \textbf{Best} & \textbf{Case 1} & \textbf{Case 2} & \textbf{Case 3} \\
\hline
Grid size & 3200$\times$128 & 80025$\times$1 & 80025$\times$1 & 25$\times$1 \\
Time (ms) & 9.84 & 11.7 & 20.2 & 217 \\
Occupancy (\%) & 37.5 & 43.75 & 25 & 25 \\
Registers/thread & 80 & 72 & 114 & 112 \\
Local Mem (GB) & 0.34 & 0.44 & 0.51 & 0.51 \\
Dynamic Shared Memory (bytes) & 192 & 0 & 0 & 0 \\
Latency ($\mu$s) & 5.865 & 6.239 & 6.246 & 6.985 \\
\hline
\end{tabular}
\end{table}
Table~\ref{tab:HLL_OMP_variants} quantifies OpenMP directive sensitivity for the \texttt{HLL()} kernel. Alternative formulations vary runtime by 22$\times$ through grid serialization and increased spilling. Only the best-performing variant allocates shared memory (192~bytes), reducing local-memory traffic. Directive changes inhibit this optimization. A final example of OpenMP rigidity emerges when applying \texttt{collapse} to the \texttt{Reconstruct()} kernel. This kernel contains several GPU-managed array wrappers (\texttt{Ary1D}, \texttt{Ary2D}) and local matrices with macro-defined dimensions that are not always statically known to the compiler. These features prevent NVHPC 24.5 from proving iteration independence across all three loops. Both \texttt{collapse(3)} and \texttt{teams distribute parallel for} fail as a result. Simplifying these abstractions into plain C-style pointer arrays can restore parallelization and reduce spilling. However, this requires intrusive refactoring and breaks one-to-one correspondence with the OpenACC version. In summary, high parallelism in the 3D Orszag–Tang problem enables both models to hide spilling latency. In 2D, limited concurrency combined with C++ abstraction overhead causes OpenMP performance breakdown. This confirms performance portability between OpenACC and OpenMP is strongly kernel-size-dependent, with small-to-medium kernels constituting the most challenging regime for OpenMP GPU offloading.

\subsection{\gpluto{} Performance on LUMI (AMD MI250X)}
\label{sec:orszagtangLumi}
Table~\ref{tab:lumi_kernels} reports MI250X obtained results. One can suddenly note that performance differs from A100, despite identical code. 
\begin{table}[H]
\centering
\scriptsize
\caption{\gpluto{} kernel metrics on AMD MI250X (\texttt{ROCm} 6.2.4), 3D Orszag-Tang. Metrics reflect single-GCD execution; aggregate MI250X peak (220 CUs, 3.2~TB/s) represents the full MCM capability. HLLD$_i$() benefits from unit-stride access; other kernels remain latency-bound.}
\label{tab:lumi_kernels}
\begin{tabular}{|l|c|c|c|c|c|c|}
\hline
\textbf{Metric} & \textbf{\texttt{HLLD$_i$()}} & \textbf{\texttt{HLLD$_{j,k}$()}} & \textbf{\texttt{Rec()}} & \textbf{\texttt{RHS()}} & \textbf{\texttt{CT\_EMF()}} & \textbf{\texttt{US()}} \\
\hline
Time (ms) & 29.7 & 110.6 & 25.7 & 35.0 & 32.4 & 119.8 \\
Runtime fraction (\%) & 1.9 & 13.8 & 4.8 & 6.6 & 12.1 & 22.4 \\
FP64 ThrP (GFlop/s) & 1156 & 312 & 3584 & 68 & 33 & 3 \\
Mem BW (GB/s) & 566 & 248 & 417 & 447 & 717 & 290 \\
AI (flop/byte) & 2.04 & 1.26 & 8.60 & 0.15 & 0.04 & 0.01 \\
Occupancy (\%) & 24.7 & 24.7 & 46.8 & 51.6 & 51.8 & 71.1 \\
IPC & 0.18 & 0.05 & 0.74 & 0.06 & 0.02 & 0.07 \\
L1 Hit (\%) & 75 & 52 & 85 & 69 & 24 & 37 \\
L2 Hit (\%) & 29 & 43 & 25 & 37 & 5 & 15 \\
VGPR/wavefront & 120 & 108 & 92 & 88 & 88 & 56 \\
Scratch Alloc. (bytes) & 32 & 32 & 32 & 32 & 32 & 32 \\
VALU Utilization (\%) & 15 & 4 & 58 & 3 & 0.3 & 4 \\
VMEM Utilization (\%) & 1.03 & 0.28 & 1.51 & 0.57 & 0.22 & 0.19 \\
Branch Utilization (\%) & 0.64 & 0.17 & 4.53 & 0.31 & 0.02 & 0.75 \\
\hline
\end{tabular}
\end{table}
Most kernels sustain substantial HBM traffic (up to 717 GB/s) but remain far below the 1.6 TB/s MI250X peak. Low VALU utilization ($<5\%$) and IPC ($<0.1$) for memory-intensive kernels confirm latency-limited performance rather than bandwidth saturation. Only \texttt{Reconstruct()} achieves higher efficiency (IPC 0.74, VALU 58\%, AI 8.6). L1 hit rates vary substantially: \texttt{Reconstruct()} $\approx 85\%$, \texttt{CT\_EMF()} and \texttt{US()} below 40\%; L2 rates remain modest. Irregular/strided access patterns amplify memory latency, limiting cache hierarchy effectiveness. \texttt{HLLD()} exhibits 3.7$\times$ performance gap between unit-stride ($i$-sweep: 29.7 ms) and strided sweeps ($j,k$: 110.6~ms). This correlates with L1 hit rates (75\% vs. 52\%) and VALU utilization (15\% vs. 4\%). The array layout causes $j,k$ directions to bypass L1, so increased VMEM latency causes wavefronts stall. On A100, the same anisotropy is mild (approximately 20\%), while on MI250X it dominates performance. This can be seen as a confirmation of its latency-bound behavior: despite substantial HBM traffic relative to the 1.6~TB/s per‑GCD peak on MI250X, performance does not improve under strided patterns. \texttt{RHS()} shows 33\% directional variation (vs. 14\% on A100). Achieved occupancy spans 25–71\%, yet higher occupancy does not improve performance: \texttt{US()} achieves 71\% occupancy, while \texttt{RHS()} and \texttt{CT\_EMF()} reach 50\%, but all exhibit low IPC and VALU utilization. Occupancy alone cannot hide latency under irregular access patterns. Extremely low AI and VALU utilization are observed. Execution is dominated by stalled memory operations rather than sustained data streaming. 
Unlike the A100, MI250X kernels exhibit moderate VGPR usage (56-120) and negligible spilling, though scratch allocation is non-zero. Slower \texttt{HLLD$_{j,k}$()} uses fewer VGPRs (108) than \texttt{HLLD$_i$()} (120). Register pressure is not the limiting factor. Although the per-work-item VGPR cap is nearly identical (MI250X 256 vs. A100 255), the larger CU-level register file on CDNA2 mitigates abstraction-induced spilling, shifting the primary bottleneck to memory-access patterns and unified-memory runtime page migration effects exacerbated by irregular or strided accesses~\cite{Smith2022}.

%Unlike the A100 (Section~\ref{sec:riemann}), MI250X kernels show moderate VGPR usage (56–120). Spilling is negligible, though scratch allocation is non-zero. Slower \texttt{HLLD$_{j,k}$()} uses fewer VGPRs (108) than \texttt{HLLD$_i$()} (120). Register pressure is not the limiting factor. Although the per-work-item VGPR cap is nearly identical (MI250X 256 vs. A100 255), the significantly larger CU‑level register file on CDNA2 mitigates abstraction‑induced spilling; combined with differences in the ROCm (AOCC/Clang) backend for allocation/alias analysis, the bottleneck shifts to memory‑access patterns~\cite{Smith2022}. Finally, note that unified memory may further increase sensitivity to irregular or strided accesses due to runtime page migration effects.

\section{Discussion}
\label{sec:discussion}
\begin{figure}[H]
\centering
\tiny
\begin{tikzpicture}
\begin{axis}[
    ybar,
    width=0.95\columnwidth,
    height=0.4\columnwidth,
    bar width=9pt,
    ymin=0,
    ylabel={Runtime [s]},
    symbolic x coords={A100 ACC, A100 OMP, MI250X OMP},
    xtick=data,
    nodes near coords,
    nodes near coords align={vertical},
    nodes near coords style={anchor=west, xshift=4pt},
    ymajorgrids,
    grid style={dashed,gray!30},
    enlarge x limits=0.2,
]
\addplot coordinates {
    (A100 ACC, 13.9)
    (A100 OMP, 17.2)
    (MI250X OMP, 48.1)
};
\end{axis}
\end{tikzpicture}
\caption{Total execution time for 10 steps of 3D Orszag-Tang test.}
\label{fig:orszagtang_runtime}
\end{figure}
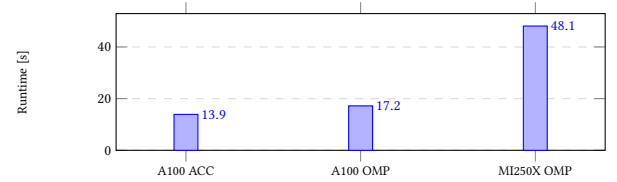

Cross-architecture analysis reveals substantial MI250X performance deterioration despite identical OpenMP code, as shown by total execution times in Figure~\ref{fig:orszagtang_runtime}. While both architectures sustain non-negligible HBM traffic, MI250X converts it into lower FP64 throughput due to poor memory coalescing under strided access. The much smaller L1/L2 cache (16~KB vector L1 per CU, 8~MB L2 vs. 192~KB L1 and 40~MB L2 on A100) amplifies non-unit-stride penalties, consistently with the low hit rates and VMEM under-utilization in Table~\ref{tab:lumi_kernels}.
The \texttt{HLLD()} 3.7$\times$ directional anisotropy on MI250X (vs. $\approx$20\% on A100) exemplifies this: strided $j,k$ sweeps bypass L1 with hit rate drops, collapsing VALU utilization and increasing VMEM stalls. Application-level performance is dictated by these latency-sensitive kernels: large FLOP/s gains ($16.5\times$ vs. A100) and high VALU utilization in \texttt{Reconstruct()} do not compensate previous latencies, resulting in a 3$\times$ slowdown shown in Figure~\ref{fig:orszagtang_runtime}, illustrating why aggregate throughput metrics obscure real bottlenecks.\footnote{Kernel-level profiling (Tables~\ref{tab:leonardo_merged} and~\ref{tab:lumi_kernels}) reports single-invocation timings under isolated conditions, while Figure~\ref{fig:orszagtang_runtime} aggregates several kernel launches per time step across spatial directions, RK stages, and auxiliary routines. Absolute times are not directly comparable.}
% Despite \texttt{Reconstruct()} achieving 16.5$\times$ higher FP64 throughput on MI250X (3584 vs. 217GFLOP/s, Tables~\ref{tab:lumi_kernels} and~\ref{tab:leonardo_merged}) and high VALU utilization (58\%), this does not translate into application speedup, with a 3$\times$ slowdown shown in Figure~\ref{fig:orszagtang_runtime} (48.1~s vs. 13.9-17.2~s on A100\footnote{Kernel-level profiling (Tables~\ref{tab:leonardo_merged} and~\ref{tab:lumi_kernels}) reports single-invocation timings under isolated conditions, while Figure~\ref{fig:orszagtang_runtime} aggregates several kernel launches per time step across spatial directions, RK stages, and auxiliary routines. Absolute times are not directly comparable.}). End-to-end performance is dictated by latency-sensitive kernels and strided-access sensitivity: \texttt{HLLD()} contributes disproportionately to the gap despite not dominating runtime, while \texttt{Reconstruct()} accounts for $< 5\%$ despite high throughput. This inversion illustrates why aggregate FLOP/s metrics obscure kernel-level bottlenecks.
Other metrics correlate with this behavior. Occupancy alone cannot hide latency without coalescing: enabled by its larger CU-level register file, low VMEM utilization and modest L2 hit rates indicate frequent wavefront stalls rather than instruction bottlenecks. Although IPC is not directly comparable across architectures, it provides a qualitative indicator that confirms this behavior. High branch efficiency on both platforms rules out divergence as a cause, confirming that memory coalescing and latency hiding drive the MI250X slowdown.
%Other metrics correlate with this behavior. Occupancy alone cannot hide latency without coalescing: despite comparable or higher occupancy on MI250X, enabled primarily by a substantially larger CU‑level register file on CDNA2, low VMEM utilization and modest L2 hits indicate frequent wavefront stalls. Although IPC is not directly comparable across architectures, it provides a qualitative indicator that the slowdown is not caused by a lack of instruction issue, but by memory stalls. Performance correlates with directionality and cache behavior rather than branch divergence (high branch efficiency on both platforms), confirming that memory coalescing and latency hiding drive the MI250X slowdown.
Table~\ref{tab:abstraction_summary} summarizes how C++ wrapper abstractions (\texttt{Ary1D}/\texttt{Ary2D}) affect performance across regimes. In 2D kernels, OpenMP's conservative pointer handling triggers massive spilling and suboptimal grid mapping (47$\times$ slowdown). In 3D, higher parallelism hides latency, recovering near-parity (6\% gap) with OpenACC through the unified CUDA backend. On MI250X, the larger CU-level register file mitigates spilling, shifting the bottleneck to memory coalescing, further amplified by the relative immaturity of OpenMP offloading on ROCm (AOCC/Clang) backend for allocation/alias analysis. Recovering A100-level performance therefore requires loop reordering, data-layout transformations, and direction-specific specialization for register management. Such interventions are well beyond straightforward \texttt{pragma} translation.
\begin{table}[H]
\centering
\footnotesize
\caption{C++ abstraction (\texttt{Ary1D}/\texttt{Ary2D}) impact on kernel behavior across architectures and programming models.}
\resizebox{\columnwidth}{!}{%
\begin{tabular}{|l|c|c|c|c|}
\hline
\textbf{Configuration} & \textbf{Register Usage} & \textbf{Spilling} & 
\textbf{Primary Bottleneck} & \textbf{Performance} \\ \hline
A100 OpenACC 2D & 104 reg/thread & None & Memory latency & Fast \\ \hline
A100 OpenMP 2D & 98 reg/thread & Massive & Spilling + grid & 
47$\times$ slower \\ \hline
A100 OpenACC 3D & 172 reg/thread & Moderate & Latency + spilling & 
Moderate \\ \hline
A100 OpenMP 3D & 182 reg/thread & Moderate & Latency (hidden) & 
6\% slower \\ \hline
MI250X OpenMP 3D & 108-120 VGPR/wf & Minimal & Strided access & 
3$\times$ slower \\ \hline
\end{tabular}
}
\label{tab:abstraction_summary}
\end{table}
%Table~\ref{tab:abstraction_summary} summarizes how implemented C++ wrapper abstractions (\texttt{Ary1D}/\texttt{Ary2D}) affect performance across regimes. In 2D low-parallelism kernels ($10^7$ cells), OpenMP's conservative pointer handling triggers massive spilling ($8.8 \times 10^7$ loads) and suboptimal grid mapping (47$\times$ slowdown). In 3D ($4.4\times 10^7$ cells), higher parallelism hides latency, achieving near-parity (6\%) despite significant spilling. On MI250X, the significantly larger CU‑level register file mitigates abstraction‑induced register pressure (the per‑work‑item VGPR cap is nearly identical: 256 vs. 255), shifting the bottleneck to memory coalescing. Because OpenACC and OpenMP achieve near-parity on A100 through the unified CUDA backend, the MI250X gap originates from architectural sensitivity to strided access amplified by OpenMP offloading maturity. Recovering A100-level performance on MI250X requires non-trivial interventions beyond straightforward pragma translation: loop reordering, data-layout transformations, and register management. 
These effects are particularly visible in smaller kernels on A100 as well (Section~\ref{sec:riemann}), where OpenMP overheads (mapping, team setup, privatization) combine with complex C++ wrappers to trigger excessive register pressure and spilling. In the 3D Orszag–Tang problem, higher parallelism hides spilling latency, achieving near-parity despite moderate register pressure in individual kernels. Notably, the observed occupancy range reflects architectural and compiler differences rather than insufficient parallelism. An alternative approach to mitigate register pressure would involve explicitly constraining register usage through compiler flags (e.g., \texttt{-maxrregcount} on NVHPC) to increase occupancy at the cost of additional spilling. However, optimal limits are architecture- and kernel-specific, undermining portability. We therefore prefer algorithmic transformations that reduce intrinsic register pressure. Finally, an additional factor is unified memory on MI250X, whose automatic page migration may amplify strided-access overhead. Isolating this contribution for a pointer-intensive C++ code like \gpluto{} requires explicit allocation strategies that underscore practical portability challenges.
%An additional factor is the use of unified memory (automatic page migration) on MI250X versus managed memory on A100. While unified memory simplifies porting, it can introduce overhead from page migration under irregular access patterns. Our kernel-level analysis indicates strided-access sensitivity and limited latency hiding as primary sources of the performance gap, though unified memory may amplify these effects. Isolating the pure architectural contribution would require explicit allocation strategies for a pointer-intensive C++ code like \gpluto{}, resulting into an engineering effort beyond this study's scope that underscores practical portability challenges. The observed gap reflects combined architectural and memory-management effects, both fundamental to real-world application porting. 

\section{Conclusions}
\label{sec:conclusions}
This work assessed performance portability of directive-based GPU programming by porting the production astrophysical code \gpluto{} to NVIDIA A100 and AMD MI250X GPUs using OpenACC and OpenMP. Directive-based models ease incremental porting of legacy codes, but require architecture-aware tuning, detailed profiling, and data layout transformations to achieve competitive performance. 

A key outcome is that performance portability is governed more by compiler maturity and backend behavior than by the programming model alone. The NVHPC toolchain benefits from over a decade of CUDA-targeted optimization heuristics, while the ROCm OpenMP implementation, though based on the same LLVM infrastructure, represents a more recent engineering effort with GPU-specific passes still under active development. On NVIDIA A100, the shared \texttt{CUDA} backend enables near-parity between OpenACC and OpenMP. In contrast, the same OpenMP implementation and workload on AMD MI250X exhibits a slowdown of $\sim 3 \times$. These are not explained by insufficient parallelism, but arise from strong sensitivity to strided memory-access patterns, limited latency hiding, and variable compiler optimization across kernels. We observe that architectural bottlenecks manifest differently across vendors: on MI250X, high sustained HBM bandwidth does not translate into throughput for irregular or directionally dependent kernels, while on NVIDIA GPUs C++ abstractions primarily manifest through register pressure and spilling. We also show that application-level performance is dictated by a small number of latency-sensitive kernels: large speedups in secondary kernels (e.g., \texttt{Reconstruct()}) do not compensate for slowdowns along the critical path. 

Overall, porting from OpenACC on NVIDIA to OpenMP on AMD is not a "search and replace" process: recovering performance requires intrusive transformations such as loop reorderings, data-layout changes, software transpositions, and direction-specific specializations. Future work should also isolate the impact of unified versus managed memory on AMD platforms and explore explicit data-movement strategies that avoid page-migration overhead while preserving maintainability. Furthermore, extending the analysis to Intel GPUs would provide a third vendor perspective, though system-dependent \texttt{unified\_shared\_memory} (USM) behavior and compiler maturity introduce non-trivial engineering challenges~\cite{Elwasif23} (\texttt{mappers}, explicit allocation APIs, compiler-specific paths). Moreover, inspecting OpenMP implementations with alternative compilers (e.g., upstream Clang/LLVM) on NVIDIA hardware to isolate the impact of NVHPC-specific optimizations from intrinsic OpenMP model characteristics would be a valuable insight. Quantifying porting effort alongside performance gain, e.g., refactoring complexity per kernel, would further make these findings more actionable for the HPC community.
%While Intel’s LLVM-based compilers document OpenMP \texttt{unified\_shared\_memory} (USM) support for Intel GPUs~\cite{OpenMP2021}, its effective availability and behavior remain implementation- and system-dependent~\cite{Elwasif23}, thus requiring additional engineering effort (\texttt{mappers}, explicit allocation APIs, compiler-specific paths). Moreover, it would be interesting, for future purposes, to inspect OpenMP implementations with alternative compilers (e.g., upstream Clang/LLVM) on NVIDIA hardware to isolate the impact of NVHPC-specific optimizations from intrinsic OpenMP model characteristics. Additionally, extending the analysis to Intel GPUs would provide a third vendor perspective on directive-based portability challenges. 
While such work increases development cost, multi-vendor support remains essential for broad access to EuroHPC systems, long-term sustainability of scientific codes, and exploitation of architecture-specific strengths. Combining metric-based approaches with detailed profiling therefore represents an important direction for future research. In summary, correctness portability across modern GPUs is readily achievable, whereas performance portability remains a significant challenge.

\begin{acks}
All authors contributed equally. The authors acknowledge M. Bettencourt and G. Rossi for their collaboration. This work was supported by the SPACE Centre of Excellence (EU Grant No.~101093441). We acknowledge ISCRA for awarding access to the LEONARDO supercomputer, owned by the EuroHPC Joint Undertaking, hosted by CINECA (Italy); and the EuroHPC Joint Undertaking for access to LUMI, hosted by CSC (Finland) and the LUMI consortium. AI assistants (ChatGPT, Claude) were used solely for light editing and proofreading; all scientific content is the original work of the authors.
\end{acks}

\bibliographystyle{ACM-Reference-Format}
\bibliography{references}

\end{document}